\documentclass[iop]{emulateapj}
\usepackage{graphicx}
\usepackage{epstopdf}
\usepackage{apjfonts}
\usepackage[usenames,dvips]{color}
\usepackage{longtable}
\usepackage{amsmath}
\usepackage{threeparttable}
\usepackage{changes}
\usepackage{lipsum}

\usepackage[bookmarks,breaklinks]{hyperref}
   \hypersetup{
     colorlinks,
     citecolor=blue,
     linkcolor=blue,
    }    
\begin{document}
%
\shorttitle{Imaging M87 with the EHT}
\shortauthors{R.-S. Lu et al.}
\title{Imaging the Supermassive Black Hole Shadow and Jet Base of M87 with the Event Horizon Telescope}
\author{
Ru-Sen \ Lu \altaffilmark{1},
Avery E.\ Broderick\altaffilmark{2,3},
Fabien \ Baron\altaffilmark{4},
John D.\ Monnier\altaffilmark{5},
Vincent L.\ Fish\altaffilmark{1},
Sheperd S.\ Doeleman\altaffilmark{1,6},
Victor \ Pankratius \altaffilmark{1}
}
\email{rslu@haystack.mit.edu}
\altaffiltext{1}{Massachusetts Institute of Technology, Haystack
 Observatory, Route 40, Westford, MA 01886, USA}
 \altaffiltext{2}{Perimeter Institute for Theoretical Physics, 31 Caroline Street North, Waterloo, ON, N2L 2Y5, Canada}
\altaffiltext{3}{Department of Physics and Astronomy, University of Waterloo, 200 University Avenue West, Waterloo, ON, N2L 3G1, Canada}
\altaffiltext{4}{Department of Physics and Astronomy, Georgia State University,  25 Park Place NE, Atlanta, GA 30303, USA}
\altaffiltext{5}{Department of Astronomy, University of Michigan, 918 Dennison Building, Ann Arbor, MI 48109-1090, USA }
\altaffiltext{6}{Harvard-Smithsonian Center for Astrophysics, 60 Garden    St., Cambridge, MA 02138, USA} 
 
\begin{abstract}
The Event Horizon Telescope (EHT) is a project to assemble a Very Long Baseline Interferometry (VLBI) network of mm wavelength dishes that can resolve strong field General Relativistic signatures near a supermassive black hole. As planned, the EHT will include enough dishes to enable imaging of the predicted black hole ``shadow'', a feature caused by severe light bending at the black hole boundary.  The center of M87, a giant elliptical galaxy, presents one of the most interesting EHT targets as
it exhibits a relativistic jet, offering the additional possibility of studying jet genesis on Schwarzschild radius scales. Fully relativistic
models of the M87 jet that fit all existing observational constraints now allow horizon-scale images to be generated. We perform realistic VLBI simulations of M87 model images to examine detectability of the black shadow with the EHT, focusing on a sequence of model images with a changing jet mass load radius. When the jet is launched close to the black hole, the shadow is clearly visible both at 230 and 345\,GHz. The EHT array with a resolution of 20--30$\mu$as resolution ($\sim$2--4 Schwarzschild radii) is able to image this feature independent of any theoretical models and we show that imaging methods used to process data from optical interferometers are applicable and effective for EHT data sets. We demonstrate that the EHT is also capable of tracing real-time
structural changes on a few Schwarzschild radii scales, such as those implicated by VHE flaring activity of M87. While inclusion of ALMA in the EHT is critical for shadow imaging, generally the array is robust against loss of a station.

\end{abstract}

\section{INTRODUCTION}
\label{introduction}

The supermassive black holes at the centers of the Milky Way (Sgr A*) and the giant elliptical galaxy M87, present the largest apparent event horizons (by nearly an order of magnitude) from the Earth. The characteristic angular scale for a black hole is the Schwarzschild radius, $R_{sch}\equiv2GM/c^2$ where $G$ is the gravitational constant and $c$ is the speed of light, resulting in an angular scale of $10~{\rm \mu as}$ for Sgr A*, where we have adopted a mass of $4.3\times10^6 M_\odot$ and a distance of $8.3~{\rm kpc}$~\citep{2008ApJ...689.1044G,2009ApJ...692.1075G}. The angular size of M87 is less clear at present, depending upon its mass which is a subject of some debate. Recent stellar dynamics observations imply a mass of $6.6\times10^9 M_\odot$~\citep{2011ApJ...729..119G}, which at an assumed distance of 17.9~Mpc implies an angular scale of $7~{\rm \mu as}$, and the value we adopt fiducially here. However, the gas-dynamical mass estimate is $3.5\times10^9 M_\odot$~\citep{2013ApJ...770...86W}, a factor of roughly two lower, implying a correspondingly smaller angular size.

Around a black hole, a dark region silhouetted against a backdrop of bright optically thin emission (known as a ``black hole shadow'') is predicted by general relativistic lensing~\citep{1973blho.conf..215B,1979A&A....75..228L,2000ApJ...528L..13F,2006ApJ...636L.109B,2006ApJ...638L..21B,2006MNRAS.367..905B}. While the emission region depends on the details of the underlying accretion process and of the emission mechanisms, the shadow feature is nearly independent of the spin or orientation of the black hole to within 10\,\%, making possible a direct test for the presence of a horizon~\citep{2009ApJ...701.1357B}.  For a non-rotating black hole, the shadow diameter is  $\sqrt{27}R_{sch}$~\citep{2006ApJ...638L..21B}. These scales are now accessible to the Event Horizon Telescope (EHT) with $\sim$20--30$\mu$as resolution ($\theta \sim \lambda/D$) and therefore allow horizon scale imaging for the first time, providing a great opportunity to improve our understanding of the physics responsible for accretion and emission in a strong gravitational field~\citep{2008Natur.455...78D,2011ApJ...727L..36F,2012Sci...338..355D, 2009ApJ...697...45B,2011ApJ...735..110B,2011ApJ...738...38B}. 

The timescale for the Keplerian motion at the innermost stable circular orbit (ISCO) around the black hole in Sgr A* ranges from 30 minutes for a nonrotating black hole to 4 minutes for prograde orbits around a maximally rotating black hole~\citep{2009ApJ...695...59D}. These time scales are much less than the typical duration of a VLBI experiment, which violates one of the basic requirements for VLBI Earth-rotation aperture synthesis imaging. However, M87's much larger black hole mass results in a  minimal time scale of a few days at the last stable orbit, making it a practical target for direct imaging on event horizon scales. Furthermore, M87 drives a powerful relativistic jet, allowing the study of jet-launching physics. 

Horizon-scale imaging promises to elucidate many persistent uncertainties in the jet formation process~\citep{2009ApJ...697.1164B}, including in particular: measuring the rate at which astrophysical jets accelerate and collimate; identifying where the
material that mass loads the jet originates; verification of strong, ordered magnetic fields; and unambiguously determining the location of the black hole relative to the larger-scale jet structures via the observation of the black hole shadow~\citep[cf.][]{2007ApJ...660..200L,2008JPhCS.131a2053W,2011Natur.477..185H,2012ApJ...745L..28A,2013ApJ...775...70H}.

In this paper we employ a class of simple models of M87 to assess the detectability of horizon-scale structure with the EHT by focusing on the comparison of images at wavelengths of 1.3\,mm and 0.87\,mm (230 GHz and 345 GHz).  Section~\ref{sect:models_methods} describes the jet models of M87, observation simulations and the imaging method. In section~\ref{sect:results}, we report our findings, discuss their implications. We summarize our conclusions in section~\ref{sect:summ}.

\section{MODELS AND METHODS}
\label{sect:models_methods}
\subsection{Models of M87}
M87 exhibits a prominent relativistic jet detected on a variety of scales and across the spectrum, from radio wavelengths to X-rays~\citep{1999ApJ...520..621B,1999Natur.401..891J,1999AJ....117.2185P,2002ApJ...564..683M,2007ApJ...660..200L,2008JPhCS.131a2053W}. Because the EHT now provides access to R$_{sch}$-scale structure for M87, modeling the jet in this source requires inclusion of strong general-relativistic effects that lead to horizon scale signatures. Here we employ a version of the jet model developed in~\citet{2009ApJ...697.1164B} that produces sub-horizon emission structure. We direct the reader to this source for details, restricting ourselves here to a summary of its salient features and motivations.

The high luminosity of AGN derive from conversion of gravitational energy to radiation in the deep potential well of central supermassive black holes~\citep{1995ARA&A..33..581K}. It is widely believed that AGN jets are the result of the extraction
of rotational energy via large-scale electromagnetic fields near supermassive black holes~\citep{2001Sci...291...84M}.  Less clear is the reservoir of rotational energy being tapped.  Most often, the spin of rapidly rotating black holes is implicated in the formation of the highly collimated, relativistic jets observed at large distances from AGN~\citep{1977MNRAS.179..433B}.  Recent
simulations have demonstrated that such objects are capable of generating the high kinetic luminosities observed~\citep{2009MNRAS.394L.126M,2011MNRAS.418L..79T}. However, a number of authors have also speculated that the
radio jet observed near the core of M87 may be due to a disk wind, which both serves to collimate the nascent jet and is illuminated by the copious particles present launched from the disk itself~\citep{1982MNRAS.199..883B,2005ApJ...620..878D}.

Regardless of the mechanism responsible for forming the jet, a canonical jet structure has emerged from jet formation simulations.
This consists of a force-free interior, where the electromagnetic energy density greatly exceeds that associated with any entrained
particles, surrounded by a collimating, magnetically dominated wind, where magnetic pressure is large in comparison to that of the gas. Supporting both is a hot, geometrically thick accretion flow, responsible for providing the currents that support and confine the
magnetic flux near the horizon.  Motivated by the dominance of the total flux by the radio jet at large radii, we model the jet in M87
solely in terms of the first of these, the fast force-free jet core. In this sense, we consider a maximally pessimistic model for imaging
the black hole silhouette, since the majority of the emission is produced in regions flowing outward, potentially relativistically.

The global structure of the magnetic field and outflow velocities are obtained from an approximate solution to the equations of
stationary, axisymmetric force-free electrodynamics, as described in detail within~\citet{2009ApJ...697.1164B}.  In addition, an asymptotic Lorentz factor, $\gamma_\infty$, is introduced by reducing the toroidal magnetic field from the expressions presented in~\citet{2009ApJ...697.1164B} by a factor of $\beta_\infty=\sqrt{1-\Gamma_\infty^{-2}}$.  This corresponds to changing the pitch angle of the magnetic field lines in the equatorial plane, and would physically result from a mass loading of the magnetic
field lines.  Here $\Gamma_{\rm \infty}=5$, consistent with the observed proper motions of the optical knots at large distances~\citep[e.g., HST-1][]{1999ApJ...520..621B}.  In practice this makes a small difference to image morphologies.

In the presence of dynamically strong magnetic fields, self-absorbed synchrotron emission provides a natural mechanism for producing the observed radio emission. This necessarily requires a population of energetic particles, which we assume arise in the form of a nonthermal, power-law distribution of leptons, supplemented with a lower energy cutoff, as described in~\citet{2009ApJ...697.1164B}. Motivated by the spectrum of M87 above 1~mm, we assume an electron energy distribution with power-law index $p=3.38$ ($N(E) \propto E^{-p}$), corresponding to an optically thin spectral index $\alpha=1.19$ by adopting a negative sign convention ($S \propto \nu^{-\alpha}$).

Existing jet formation simulations are unable to provide strong guidance on the origin of these relativistic leptons.  Variations in
the details of the particle loading models are largely responsible for the differences in horizon-resolving images published to date~\citep[cf.][]{2009ApJ...697.1164B,2012MNRAS.421.1517D}. However, in the case of M87, where typical magnetic field
strengths are of order 100--300~G, the synchrotron cooling timescale is large in comparison to the outflow time, implying that
nonthermal population is essentially conserved on the scales of interest, greatly simplifying its modeling.  We parameterize the
particle acceleration process by imposing a particle loading radius, below which the nonthermal particle density is essentially flat, and above which it is conserved.  Given both the intrinsic uncertainty in the location of the load radius, and the likely variability of the particle acceleration process, we allow this to vary. Where a fiducial value is required, we chose $4 R_{sch}$.  Since the jet both
collimates and accelerates, moving the load radius up and down modifies the visibility of black hole shadow, cast against the
emission from the counter-jet.  

\subsection{Simulated observations}
We performed simulations of EHT observations at 230 and 345\,GHz using the MAPS (MIT Array Performance Simulator) package. The simulated sky brightness distribution was first converted into the visibility plane using an FFT. 
All images (100$\times$100 pixels in size and a pixel scale of 1.8$\mu$as) were zero-padded prior to computing the Fourier transform to allow interpolation in the Fourier domain. Visibilities were then ``observed'' based on the input array geometry (site locations), antenna properties, source positions, and observation specifications (e.g., bandwidth, integration time, and scan lengths) to produce synthetic data including thermal noise. 

MAPS samples a two dimensional-patch in the {\it uv} plane over the prescribed (channelized) bandwidth and (divided) integration time and interpolates and integrates numerically the data in the frequency and time plane to generate a complex visibility~\citep{memo006}. Samples on the {\it uv} tracks were calculated for a 12\,s integration time. Typical atmospheric coherence times at 230 GHz are 10\,s, but can be as short as 2--4\,s and as long as 20\,s depending on  weather conditions at each observing site~\citep{2009ApJ...695...59D}. Weather requirement at 345\,GHz is more challenging, but most EHT sites have suitable observing conditions in winter times. At the ALMA site, for example, the measured coherence time is $\geq$16 and $\geq$8\,s for 50\% of the time at 230 and 345\,GHz, but it will be $\geq$ 53 and $\geq$25\,s for the 25\% best conditions, which is approximately equal to the night-time median coherence time~\citep{holdaway97}.

We assumed perfect phase coherence within the integration time. In practice, visibility amplitude and phase information can be measured in terms of incoherently averaged quantities and coherence losses due to atmospheric turbulence are corrected using established algorithms tailored for high frequency observations~\citep{1995AJ....109.1391R}. Amplitude calibration will be critical, as systematic errors on measured baseline flux densities reduce the imaging sensitivity. Here, we do not include flux density calibration errors, but will consider these effects in future work. We note, however,  future experiments with improved observation and calibration strategies (e.g., inclusion of amplitude calibrators, observing with paired antennas) and with enhanced capabilities in imaging algorithms (e.g., integration of closure amplitude information) are expected to largely remove systematic uncertainties in amplitude calibration. The simulations reported here are restricted to telescope elevations above 15 degrees, where calibration issues are expected to be reduced.

We have primarily assumed an overall bandwidth of 4\,GHz at 230\,GHz and 16\,GHz at 345\,GHz, but we also consider the effects of changing bandwidth on imaging in the following sections. The assumed bandwidths would in practice be 2 and 8\,GHz in each of two polarizations and correspond to bit rates of 16 and 64\,Gbit s$^{-1}$ for four-level signals sampled at the Nyquist rate. A 16\,Gbit s$^{-1}$ data-recording system has recently been demonstrated~\citep{2012arXiv1210.5961W} and a maximum data rate of 64\,Gbit s$^{-1}$ data rate is targeted for the ALMA beam former~\citep{2013arXiv1309.3519F}. For high frequency VLBI observations, these bandwidths are relatively
small (2\,\% and 5\,\% fractional bandwidth at 230 and 345 GHz), so the source structure across each of the two frequency bands were neglected.

The assumed array at 230\,GHz consisted of stations at 8 different sites: 
Hawaii, consisting of one or more of the JCMT and SMA phased together into a single aperture; 
the Arizona Radio Observatory Submillimeter Telescope (SMT) on Mount Graham; the CARMA site in California; 
the Large Millimeter Telescope (LMT) on Sierra Negra, Mexico; 
the phased Atacama Large Millimeter/submillimeter Array (ALMA); 
the Institut de Radioastronomie Millim\'{e}trique (IRAM) 30\,m telescope on Pico Veleta (PV), Spain; 
the IRAM Plateau de Bure Interferometer (PdBI), phased together as a single aperture; and the Greenland telescope (GLT). 
At 345\,GHz, we do not include the CARMA site in California. Details of existing telescopes that may serve as participating stations in the EHT array are discussed in~\citet{2009ApJ...695...59D}. 

\begin{table*}
\centering
\caption{Assumed telescope parameters\label{Table:array}.}
\begin{tabular}{ccccc}
\hline
Facility &Code&Effective diameter&SEFD$_{230\,GHz}$&SEFD$_{345\,GHz}$\\
\hline 
             &    &[m]       &       [Jy]     &[Jy]   \\
\hline
Hawaii  &H  & 23    &      4900  & 8100 \\ 
SMT   &S & 10     &     11900& 23100\\ 
CARMA&C  & 27    &      3500 &...\\
LMT      &L  &50      &     560    &13700 \\   
ALMA   & A & 85     &     110    &140 \\
PV        &V &30      &     2900   & 5200\\
PdBI     &B &37      &    1600    & 3400\\      
GLT     &G & 12     &     4700    &8100\\
\hline
\end{tabular}
\end{table*}

\begin{figure*}[ht!]
 \begin{center}
 \includegraphics[width=0.45\textwidth,clip]{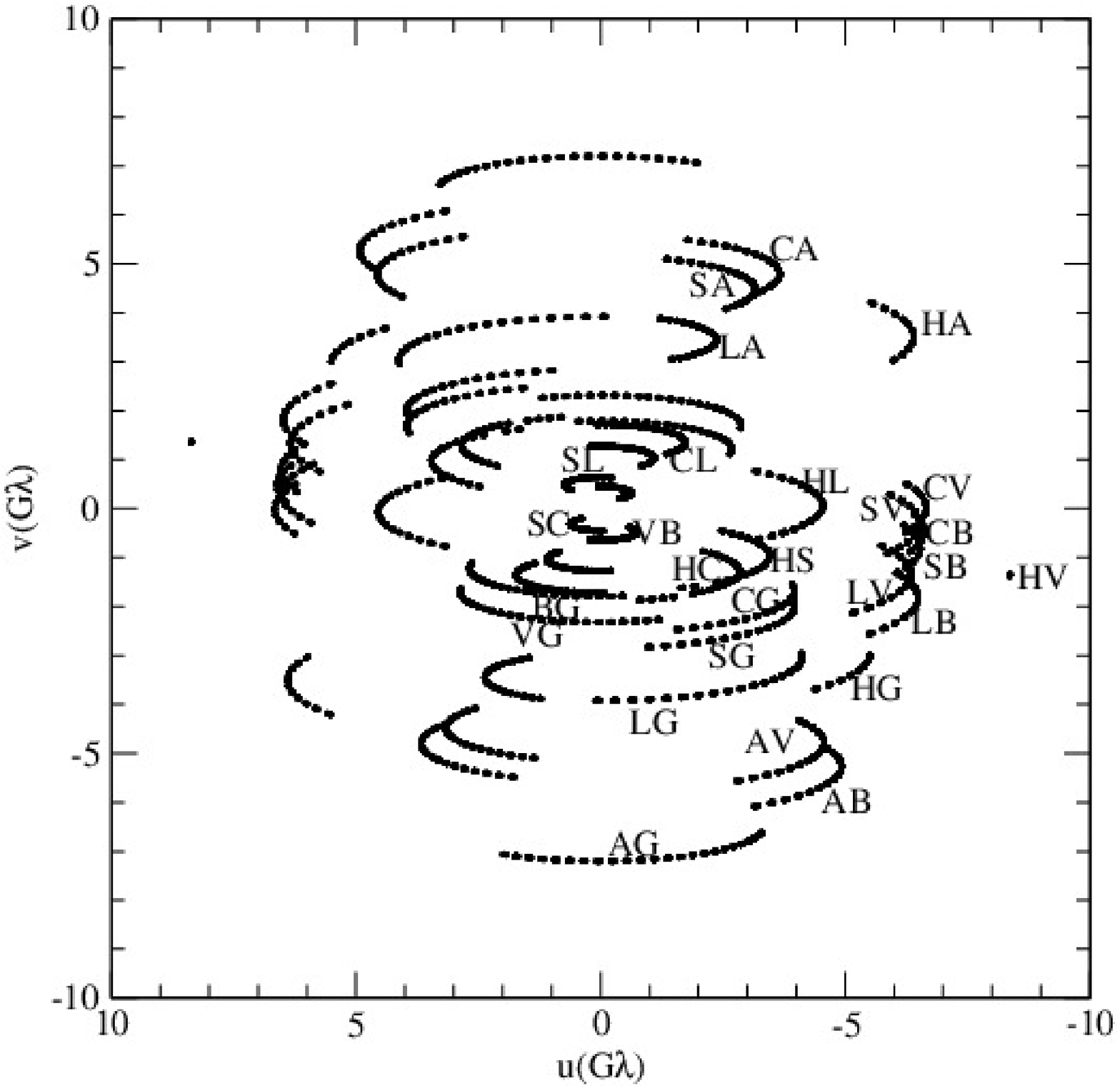}
  \includegraphics[width=0.45\textwidth,clip]{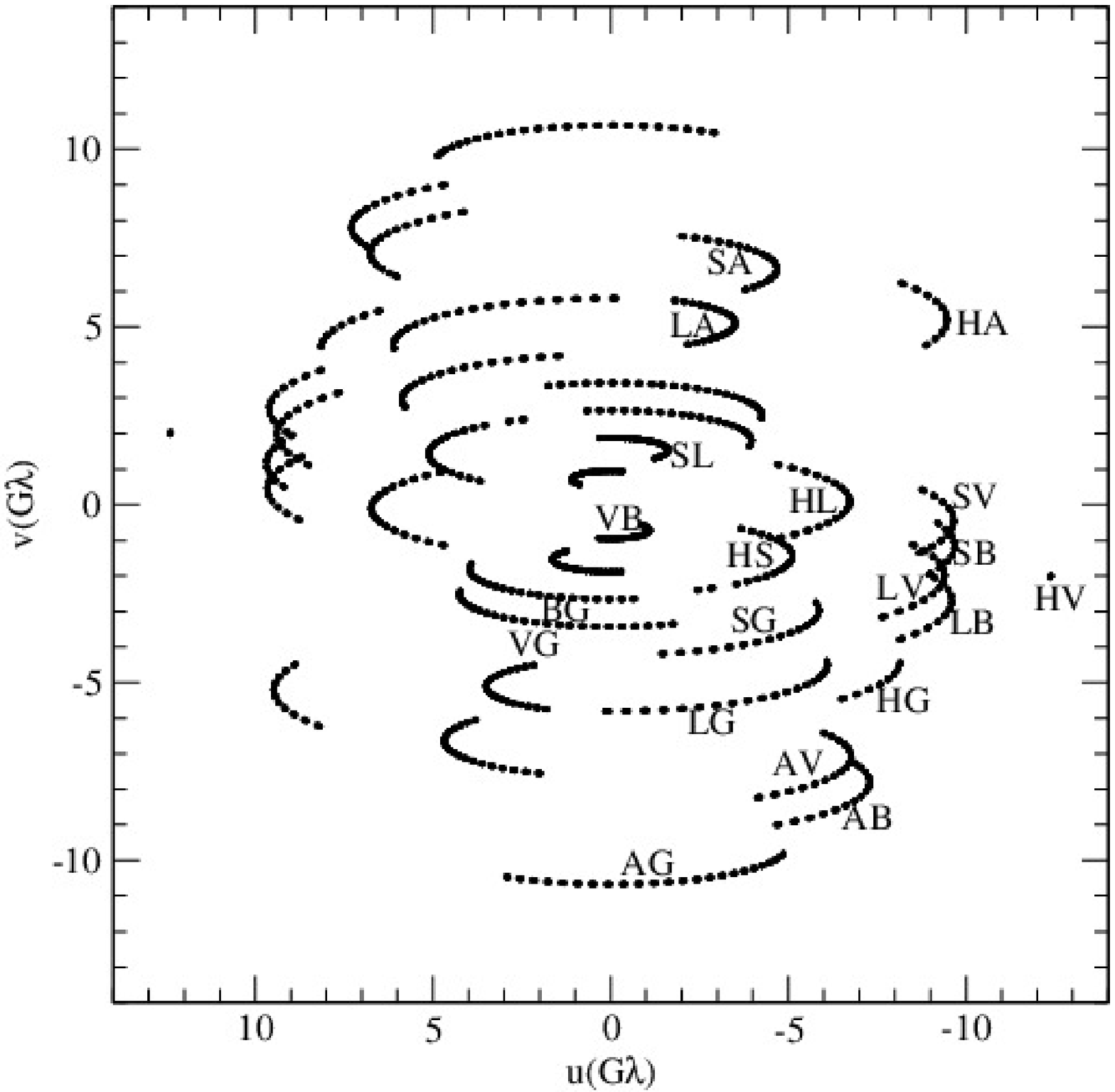}\\
 \includegraphics[width=0.45\textwidth,clip]{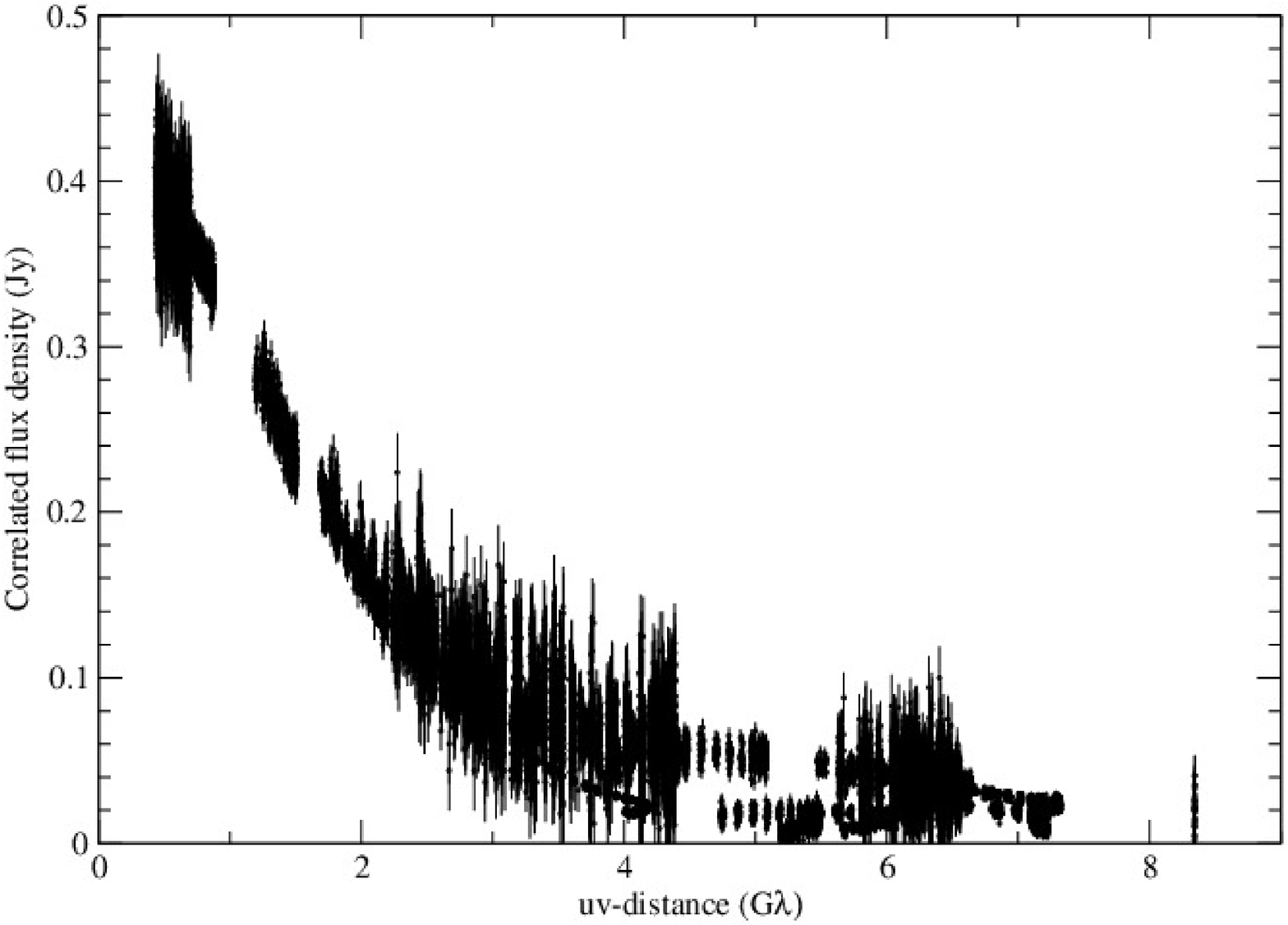}
 \includegraphics[width=0.45\textwidth,clip]{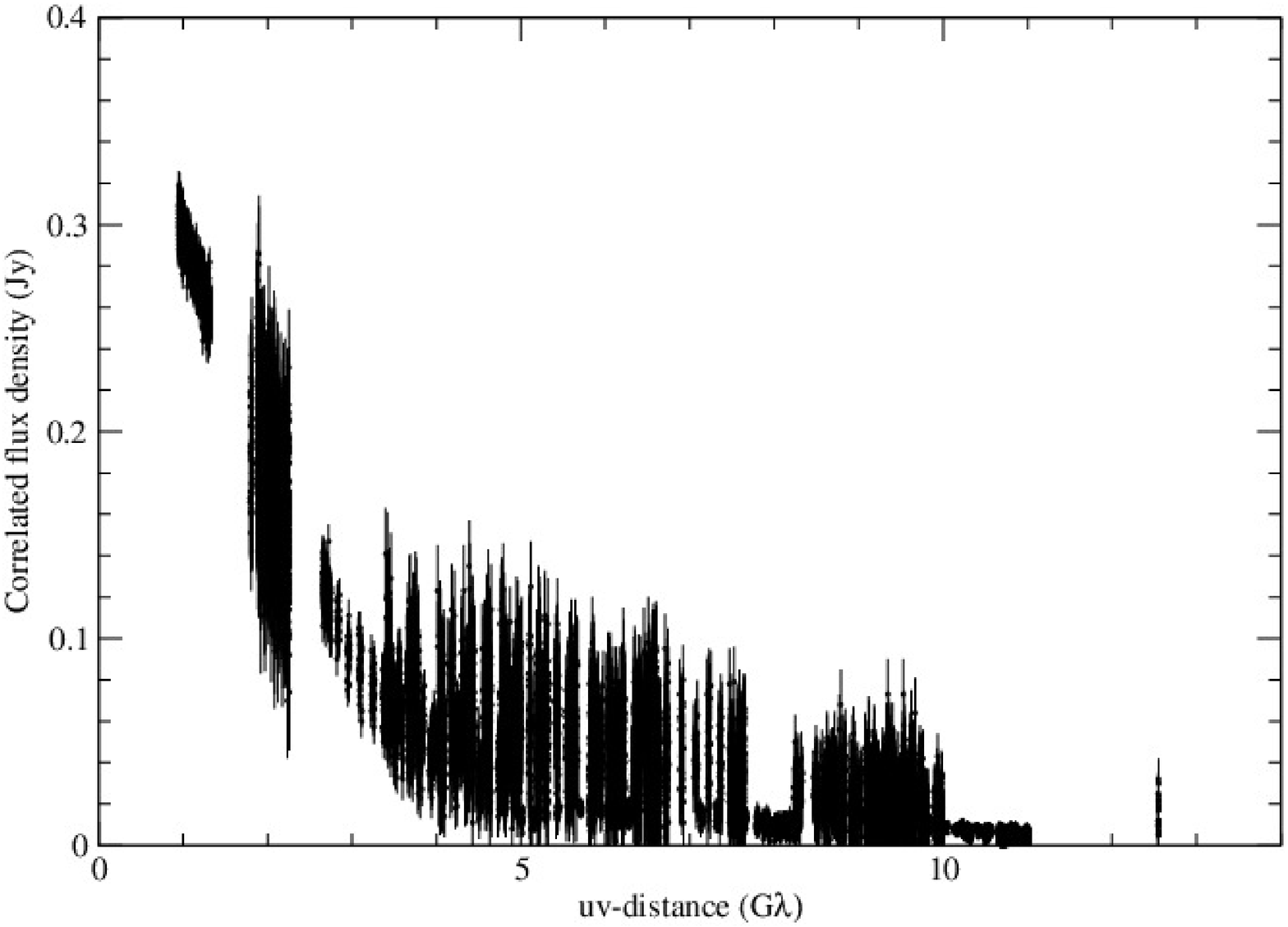}\\
 \includegraphics[width=0.45\textwidth,clip]{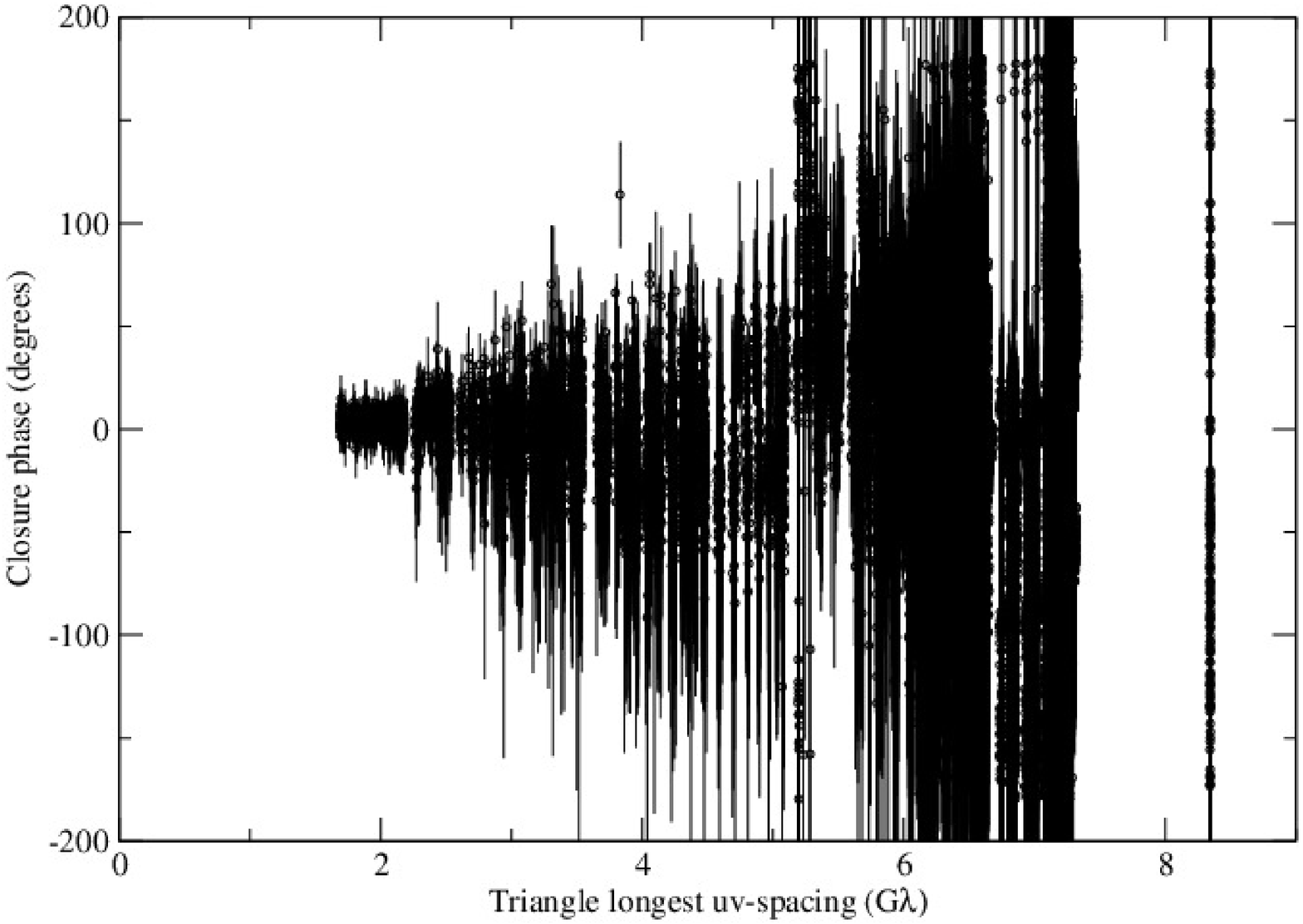}
 \includegraphics[width=0.45\textwidth,clip]{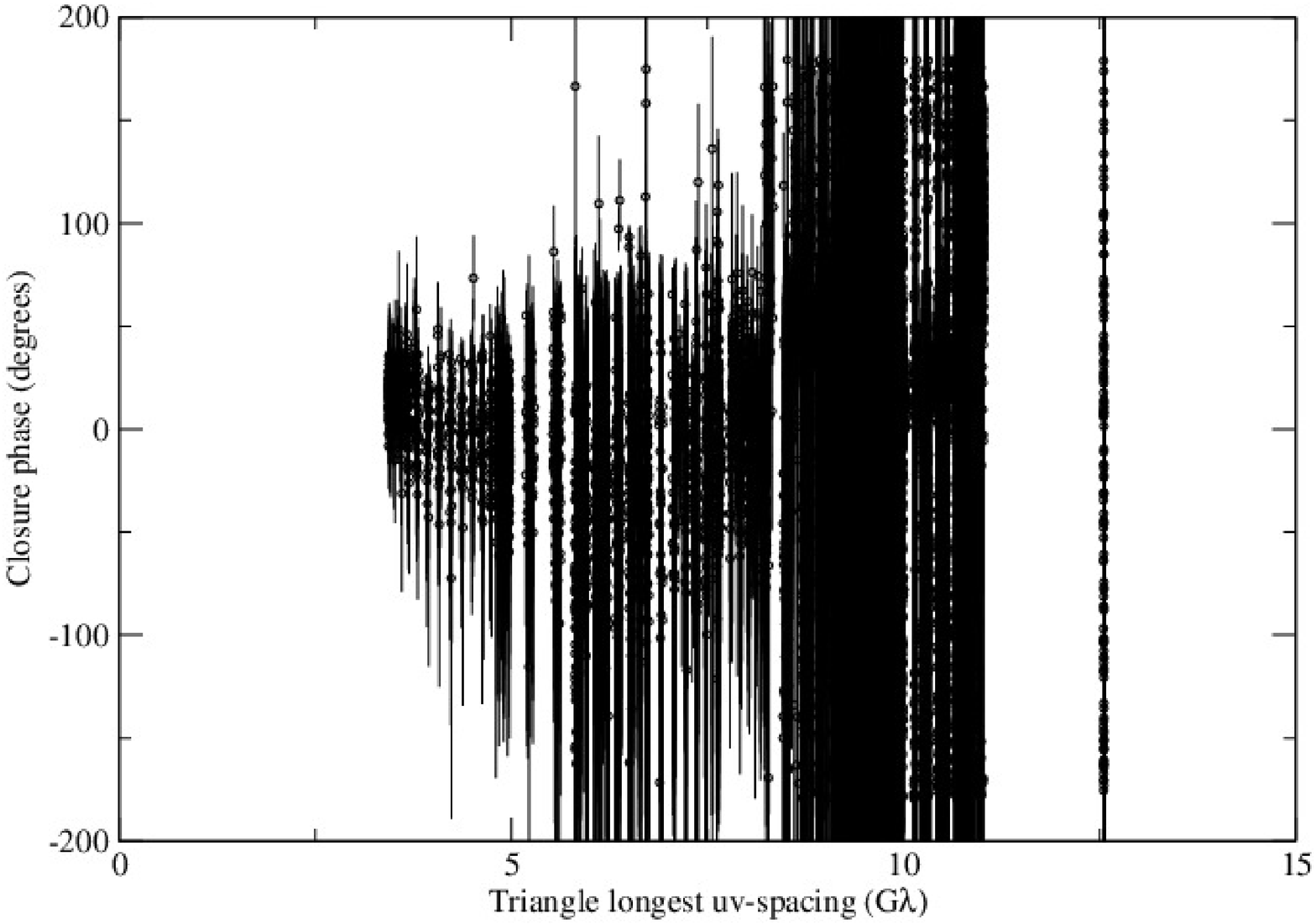}

 \caption{Top panel: {\it uv} coverage for the array used for the simulation at 230\,GHz (left) and 345\,GHz (right). Tracks are labeled by baseline (H: Hawaii; S: SMT; C: CARMA; L: LMT; A: ALMA; V: Pico Veleta; B: Plateau de Bure Interferometer and G: Greenland Telescope). At 345\,GHz, the CARMA site was not used. Middle panel: Example plot of correlated flux density as a function of {\it uv}-distance for the first image in Figure~\ref{Fig:230GHz_4G} at 230\,GHz (left) and for the first image in Figure~\ref{Fig:345GHz_16G} at 345\,GHz (right). Bottom panel: Closure phase plotted as a function of longest baseline in a triplet of stations for the same image as in the middle panel at 230\,GHz (left) and 345\,GHz (right)}
 \label{Fig:visplt}
 \end{center}
 \end{figure*}

The assumed array parameters are shown in Table~\ref{Table:array}. System equivalent flux density (SEFD) at each site was adopted and updated from~\citet{2009ApJ...695...59D} and \citet{2013arXiv1309.3519F}. Sensitivity values for each site include realistic considerations of the system temperature and its atmospheric contribution based on typical weather conditions. For phased arrays, a phasing efficiency of 90\,\% was assumed. The updates on SEFDs of the LMT and ALMA at 230\,GHz were mainly based on improved understanding of the system temperature and aperture efficiency at the LMT, and assumed atmospheric conditions at ALMA. The system sensitivity of the GLT was determined based on predicted aperture efficiencies of 0.66 and 0.7 and system temperatures of 220 and 140 K at 230 and 345\,GHz, respectively (K. Asada, private communication). Since the atmospheric contribution is highly weather-dependent, actual observations may achieve significantly different SEFDs. 
Figure~\ref{Fig:visplt}  (upper panel) shows the array {\it uv} coverage for M87. We also show a typical plot of correlated flux density (middle panel)  and closure phase (the sum of three visibility phases around a closed triangle of baselines, bottom panel) as a function of {\it uv}-distance (closure phase plotted against the longest baseline for a given triplet of baselines) at 230 and 345\,GHz, respectively. The results of the simulations are then converted into OIFITS format~\citep{2005PASP..117.1255P} and are imported into image reconstruction software for imaging\footnote{Data are available upon request to the authors.}.

\subsection{Imaging analysis}
Image reconstruction is an ill-posed inverse problem, which in general
does not admit a unique solution~\citep{2001isra.book.....T}.
To cope with the sparsity of the uv-sampling, CLEAN~\citep[e.g.,][]{1974A&AS...15..417H} and Multi-Scale-CLEAN~\citep[e.g.,][]{2008ISTSP...2..793C,2009AJ....137.4718G} algorithms are ``de-facto'' standard imaging techniques widely used in VLBI at lower frequencies. For mm-VLBI, reconstruction is complicated by the strong atmospheric corruption of the visibility phases, similar to the situation in optical interferometry.

Fortunately, practical algorithms have been developed for optical interferometry, and as such they were optimized for sparsely sampled data similar to those that the EHT will produce~\citep{2012A&ARv..20...53B}. The majority of these algorithms are based on the regularized maximum likelihood paradigm. Newer methods based on more recent signal analysis theories such as compressed sensing are in development~\citep[e.g.,][]{2010MNRAS.402.2626W,2012MNRAS.426.1223C} but have yet to demonstrate superior imaging capabilities compared to conventional regularized maximum likelihood.

\subsubsection{Image reconstruction}

Imaging algorithms based on regularized maximum likelihood attempt to find the image most compatible with the data while still constraining it to keep certain desirable or expected properties by using regularizers (such as positivity). This introduction of prior knowledge about the target/source should be as non-committal as possible, so as not to bias the image. This regularization also alleviates the ill-posed nature of the problem, by preventing convergence to bad local minima in the $\chi^2$.
However the optimal compromise between lowering the chi-squared statistic and achieving optimal regularization is non-trivial to achieve~\citep{2011A&A...533A..64R}.

Synthetic tests as well as software competitions in optical interferometry have been pitting algorithms against one another for a decade~\citep{2010SPIE.7734E..83M,2012SPIE.8445E..1EB}, demonstrating that algorithms based on similar minimization techniques achieve very close results. Therefore we selected two software algorithms representative of two main methods: the
gradient-based methods~\citep[line search or trust region, e.g.,][]{2008ISTSP...2..767L}, and the Markov Chain Monte Carlo (MCMC) method~\citep[e.g.,][]{2006ApJS..162..401S}.

BiSpectrum Maximum Entropy Method \citep[BSMEM,][]{1994IAUS..158...91B} uses a gradient descent algorithm for maximizing the posterior probability of an image, and is the last of a long line of software based on maximum entropy regularization~\citep{1984MNRAS.211..111S}. SQUEEZE~\citep{2010SPIE.7734E..78B,2012SPIE.8445E..1DB} is based on the MCMC parallel tempering algorithm to optimally sample the posterior probability distribution of the image, and is an evolution of the earlier software MACIM~\citep{2006SPIE.6268E..58I}. Contrary to BSMEM, SQUEEZE is not limited in its choice of regularizers or other prior constraints, and its algorithm is more resilient to local minima.

As part of the image reconstruction, we incorporated prior knowledge of the model total flux density. In practice, the zero-spacing flux of the VLBI-scale structure can be constrained in principle by low frequency observations via a scaling function extrapolation. Except for the total flux density, a non-informative image prior is otherwise used in all cases.

\begin{figure*}[ht!]
 \begin{center}
 \includegraphics[angle=-90,width=0.95\textwidth,clip]{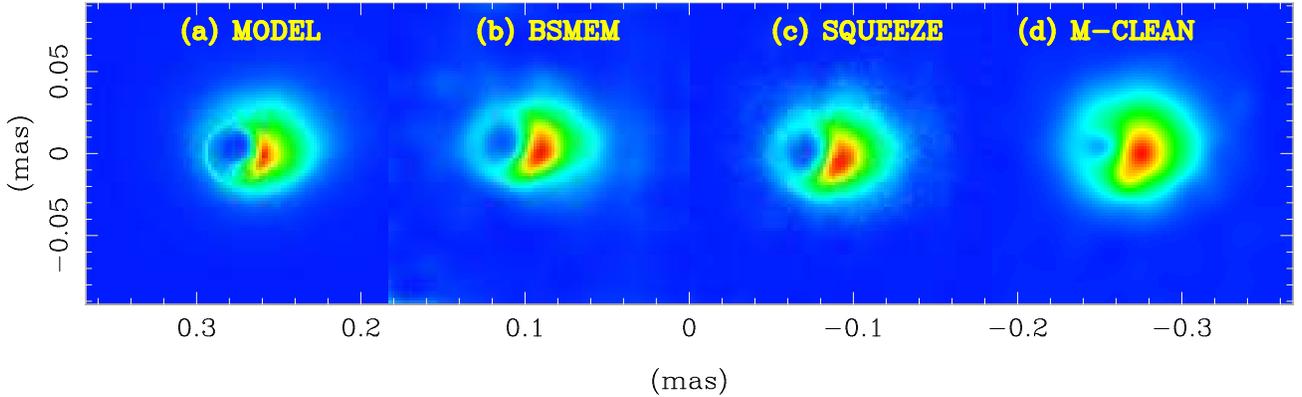}
 \caption{Comparison of imaging algorithms. A model image (the first image in Figure~\ref{Fig:230GHz_4G}) at 230\,GHz (a) and its image reconstructions with BSMEM (b), SQUEEZE (c) and Multi-Scale Clean (d). The data are shown in Figure~\ref{Fig:visplt} (left panels). The multi-scale-clean image has been restored with a circular Gaussian beam of FWHM = 15$\mu$as.}
 \label{Fig:algorithm}
 \end{center}
 \end{figure*}

In Figure~\ref{Fig:algorithm}, we show exemplary imaging reconstructions for a model image of M87 at 230\,GHz with BSMEM, SQUEEZE, and Multi-Scale-CLEAN as implemented in AIPS (Astronomical Image Processing System)~\citep{2009AJ....137.4718G}. 
The input to the BSMEM and SQUEEZE are squared visibilities and triple products/bispectrum (whose argument is the closure phase), while Multi-Scale-CLEAN uses complex visibilities. Note that in an array with N stations, N(N-1)/2 spatial components will be sampled, but phase closure only yields (N-1)(N-2)/2 independent phase estimates. Figure~\ref{Fig:algorithm} shows that algorithms based on regularized maximum likelihood are very suitable for imaging horizon scale structures using data sets (like those from the EHT) that contain only visibility amplitudes and closure phases.

\subsubsection{Image fidelity assessment}
We employ two widely used image quality metrics to evaluate the reconstructions: the mean square error (MSE) and structural dissimilarity (DSSIM) index. The MSE is a pixel-to-pixel comparison metric and is computed by averaging the squared intensity difference of the truth image and the reconstructed image pixels following

\begin{equation}
MSE =\frac{\sum\limits_{i=1}^{L}|I_i-K_i|^2}{\sum\limits_{i=1}^{L} |I_i|^2},
\end{equation}
where I$_{i}$ and K$_{i}$ are the original and reconstructed images respectively, each having L pixels.

DSSIM~\citep{Loza09} is a human visual perception-based measure and derived from structural similarity (SSIM) index, as first described in~\citet{Wang04}. The SSIM between two images I and K are defined as
\begin{equation}
\begin{array} 
{lcl} SSIM(I,K) &=& (\frac{2\mu_{I}\mu_{K}}{\mu_{I}^2+\mu_{K}^2})(\frac{2\sigma_{I}\sigma_{K}}{\sigma_{I}^2+\sigma_{K}^2})(\frac{\sigma_{IK}}{\sigma_{I}\sigma_{K}}) \\ &= & l(I,K)c(I,K)s(I,K),
\end{array}
\end{equation}
here $\mu$ is the sample mean
\begin{equation}
 \mu_I = \frac{1}{L}\sum\limits_{i=1}^{L}I_i,
\end{equation}
$\sigma$ denotes the sample standard deviation
\begin{equation}
 \sigma_I = \sqrt{\frac{1}{L-1}\sum\limits_{i=1}^{L}(I_i-\mu_I)^2},
\end{equation}
and $\sigma_{IK}$ the sample covariance
\begin{equation}
\sigma_{IK} =\frac{1}{L-1}\sum\limits_{i=1}^{L}(I_i-\mu_I)(K_i-\mu_K).
\end{equation}

These estimators are defined identically for images I and K each having L pixels and the measurement takes into account three comparisons: luminance comparison ($l(I,K)$), contrast comparison ($c(I,K)$) and structure comparison ($s(I,K)$).

As proposed in~\citet{Wang04}, the image statistics are computed locally by means of a sliding window.
The DSSIM is then $\frac{1}{\mid SSIM \mid}-1$. Before applying these metrics, all reconstructed images are aligned with their corresponding truth image using features in the images for cross-correlation. 

\section{RESULTS AND DISCUSSION}
\label{sect:results}
\subsection{Jet morphology}
  
\begin{figure}[ht!]
 \begin{center}
  \includegraphics[width=0.425\textwidth,clip]{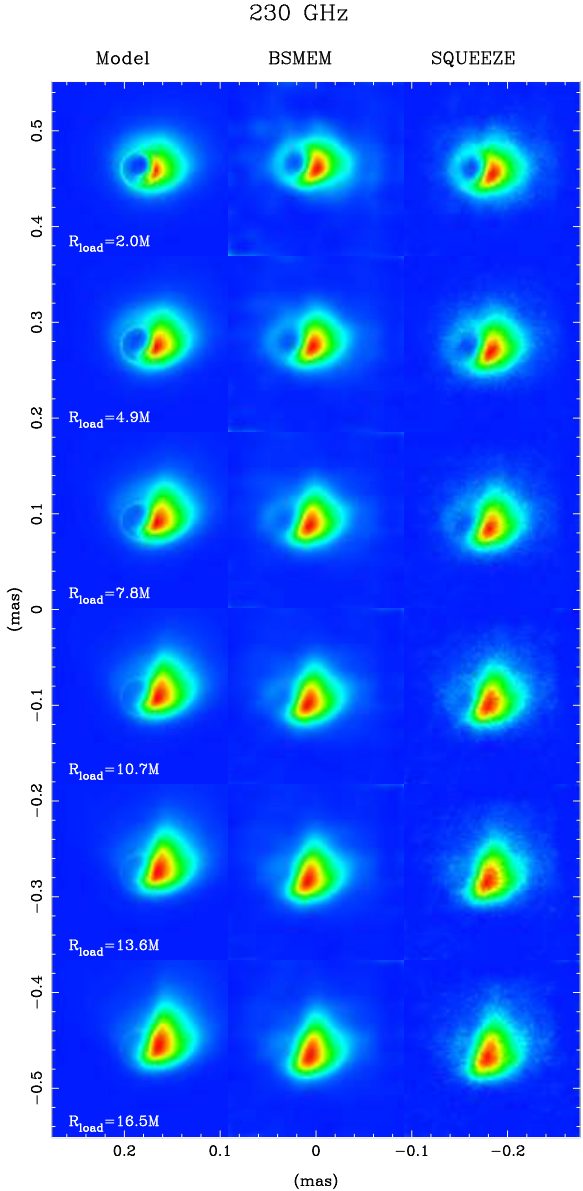}
\caption{Sequential model images of M87 at 230\,GHz and their corresponding reconstructions. The left column shows the model images (with increasing jet load radius from top to bottom). Reconstructed images are shown in the middle column using BSMEM and in the right column using SQUEEZE. The assumed bandwidth is 4\,GHz. The model images are aligned on black hole coordinates, while reconstructed images are aligned on emission centroid.}
 \label{Fig:230GHz_4G}
 \end{center}
 \end{figure}

\begin{figure}[ht!]
 \begin{center}
 \includegraphics[width=0.425\textwidth,clip]{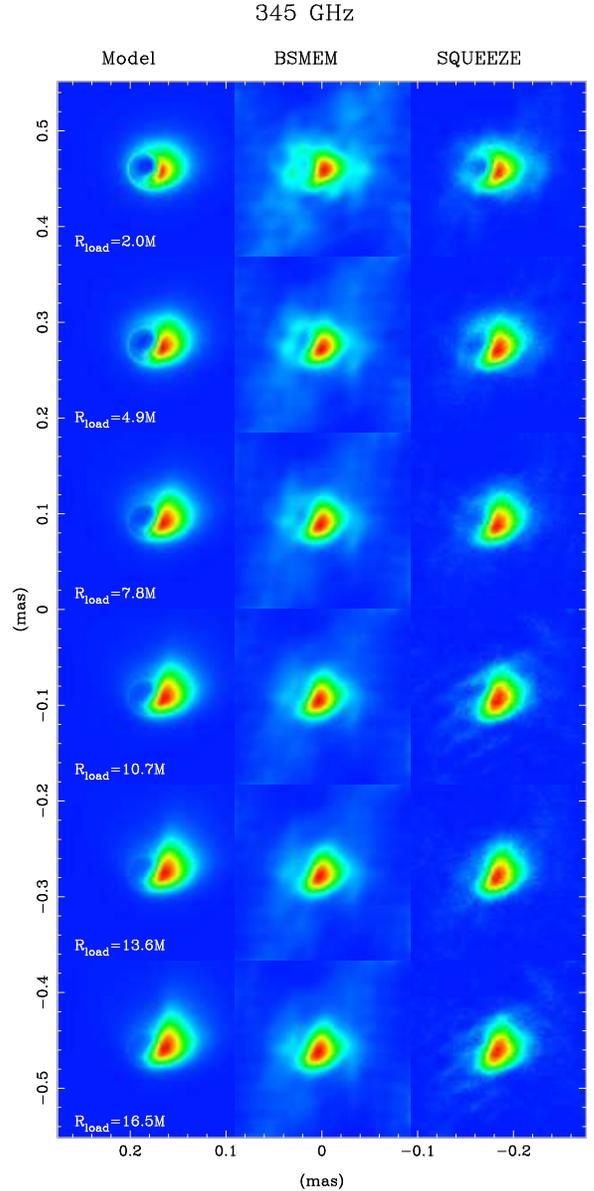}
 \caption{Same as Figure~\ref{Fig:230GHz_4G}, but for 345\,GHz and a bandwidth of 16\,GHz.}
 \label{Fig:345GHz_16G}
 \end{center}
 \end{figure}

In Figure~\ref{Fig:230GHz_4G} (left column), we show a sequence of images of the M87 jet at 230\,GHz as the load radius increases in distance (at intervals of $\sim$ 2.9\,M\footnote{When units with G = c = 1 are used, 1 M is  $\frac{R_{sch}}{2}$ and is $\sim 10^{15}$ cm for the adopted mass.} starting from $\sim$ 2M) from the black hole, leading to a gradual change in emission structure. Also shown are the reconstructed images using BSMEM (middle column) and SQUEEZE (right column)\footnote{An animation of the model images and their SQUEEZE reconstructions is available in the online version of the journal.}. 
The model intensity distributions at 345\,GHz and their reconstructions are shown in Figure~\ref{Fig:345GHz_16G}.
Table~\ref{Table:quality230} and ~\ref{Table:quality345} show the measure of quality metrics of MSE and DSSIM 
for the reconstructed images at 230 and 345\,GHz.  At 230\,GHz, both of these two metrics indicate that the model images are
fairly well reconstructed. The BSMEM reconstructions at 345\,GHz are not as good as those by SQUEEZE, with a large fraction of the flux spread over the whole reconstructed image. From both of these metrics, the reconstructions by SQUEEZE are all better than those by BSMEM, indicating that the former, which has more stringent requirement on flux location, is more suitable for imaging reconstructions for our considerations. 

In all cases, the jet emission is asymmetric and the most luminous part of the emission is due to relativistic boosting and aberration effects when the helically moving outflow approaches the observer. As the jet load radius decreases, the shadow cast by the black hole becomes more visible  (below $\sim$11\,M, Figures~\ref{Fig:230GHz_4G}--\ref{Fig:345GHz_16G}). This is because the emission from the counter jet, which acts as the back-light, has not been effectively de-boosted and is lensed by the black hole. As the jet launches gradually farther from the black hole, the counter jet gets weaker due to the beaming effect and the shadow becomes invisible~\citep{2009ApJ...697.1164B}. The structure seen at 345 GHz is very similar to that at 230\,GHz, but the effective emitting region around the shadow is smaller due to the decrease in optical depth. The emission peak is $\sim$ 4--6 M and 4--5 M from the black hole at 230 and 345\,GHz. 

The sizes of the emitting region are in the range of 40--45\,$\mu$as at 230\,GHz and 23--25\,$\mu$as at 345\,GHz if they are modeled as a circular Gaussian from data on baselines between CARMA, SMT and Hawaii alone. These sizes at 230\,GHz are very consistent with the recent measurements using the same array~\citep{2012Sci...338..355D}. Time-dependent GRMHD models also predict very compact Gaussian sizes in the range of 33--44\,$\mu$as for a $6.6\times10^9\,M_\sun$ black hole~\citep{2012MNRAS.421.1517D}. It should be emphasized that although current observations only allow
comparison of geometric models, observations with the full EHT array will enable imaging of the shadow feature and more detailed tests of physically predicted models.

The sequential launching of jets at larger distance from the black hole over time also serves as a test case for 
monitoring of the jet during flaring activity. It is known that the ejection of a new relativistic jet component on 100 $R_{sch}$ scales is connected with VHE $\gamma$-ray flares on a time scale of days~\citep{2009Sci...325..444A}. However, the black hole vicinity on a few $R_{sch}$ scales is only visible at mm/sub-mm wavelengths due to the opacity effects and only with the VLBI technique due to the required resolution. On these scales, the time scale for structural changes in M87 is comparable to the time scale of the observed VHE activity. Figure~\ref{Fig:230GHz_4G} and \ref{Fig:345GHz_16G} demonstrate the unique ability of EHT to trace the structural changes on a few $R_{sch}$ scales, which can be obtained by real time monitoring during VHE flare activities.

\begin{table}
\centering
\caption{Quality assessment with MSE (mean square error) and DSSIM (structural dissimilarity) for images at 230\,GHz. For both metrics, lower values indicate better reconstruction quality\label{Table:quality230}.}
\begin{tabular}{ccc|cc}
\hline
&\multicolumn{2}{ c| }{MSE}&\multicolumn{2}{ c }{DSSIM}\\
\cline{2-3}
\cline{4-5}
\# &BSMEM&SQUEEZE&BSMEM&SQUEEZE\\
\hline
01 &  0.059 &  0.019 &  0.632 &  0.103 \\
02 &  0.026 &  0.009 &  0.341 &  0.060 \\
03 &  0.017 &  0.007 &  0.263 &  0.066 \\
04&  0.013 &  0.007 &  0.216 &  0.077 \\
05&  0.011 &  0.007 &  0.203 &  0.092 \\
06&  0.010 &  0.008 &  0.196 &  0.092 \\
\hline
\end{tabular}
\end{table}

\begin{table}
\centering
\caption{Quality assessment with MSE and DSSIM for images at 345\,GHz. For both metrics, lower values indicate better reconstruction quality\label{Table:quality345}.}
\begin{tabular}{ccc|cc}
\hline
&\multicolumn{2}{ c| }{MSE}&\multicolumn{2}{ c }{DSSIM}\\
\cline{2-3}
\cline{4-5}
\#  &BSMEM&SQUEEZE&BSMEM&SQUEEZE\\
\hline
01&  0.367 &     0.038&  3.210 &0.180\\ 
02&  0.270 &   0.012  &  2.001 &  0.111 \\ 
03&  0.202 &   0.016  &  1.420 &  0.181 \\ 
04&  0.157 &   0.011  &  1.100 &  0.182 \\ 
05&  0.145 &   0.015  &  1.033 &  0.229 \\ 
06&  0.154 &   0.025  &  1.080 &  0.339 \\  
\hline
\end{tabular}
\end{table}

\subsection{Observational uncertainties and limitations}
In order to examine the black hole shadow detectability in M87, we have assumed a thermal-noise limited VLBI array at 230 and 345\,GHz. However, real VLBI observations may face more stringent limitations, resulting in a decrease in the array {\it uv}-coverage and sensitivity. Furthermore, VLBI observations at short millimeter wavelengths (especially shorter than 1.3\,mm) are currently still under development and the actual observations may achieve significantly different values of the SEFD from what we have assumed at each site.

\begin{figure*}[ht!]
 \begin{center}
\includegraphics[angle=-90,width=0.95\textwidth,clip]{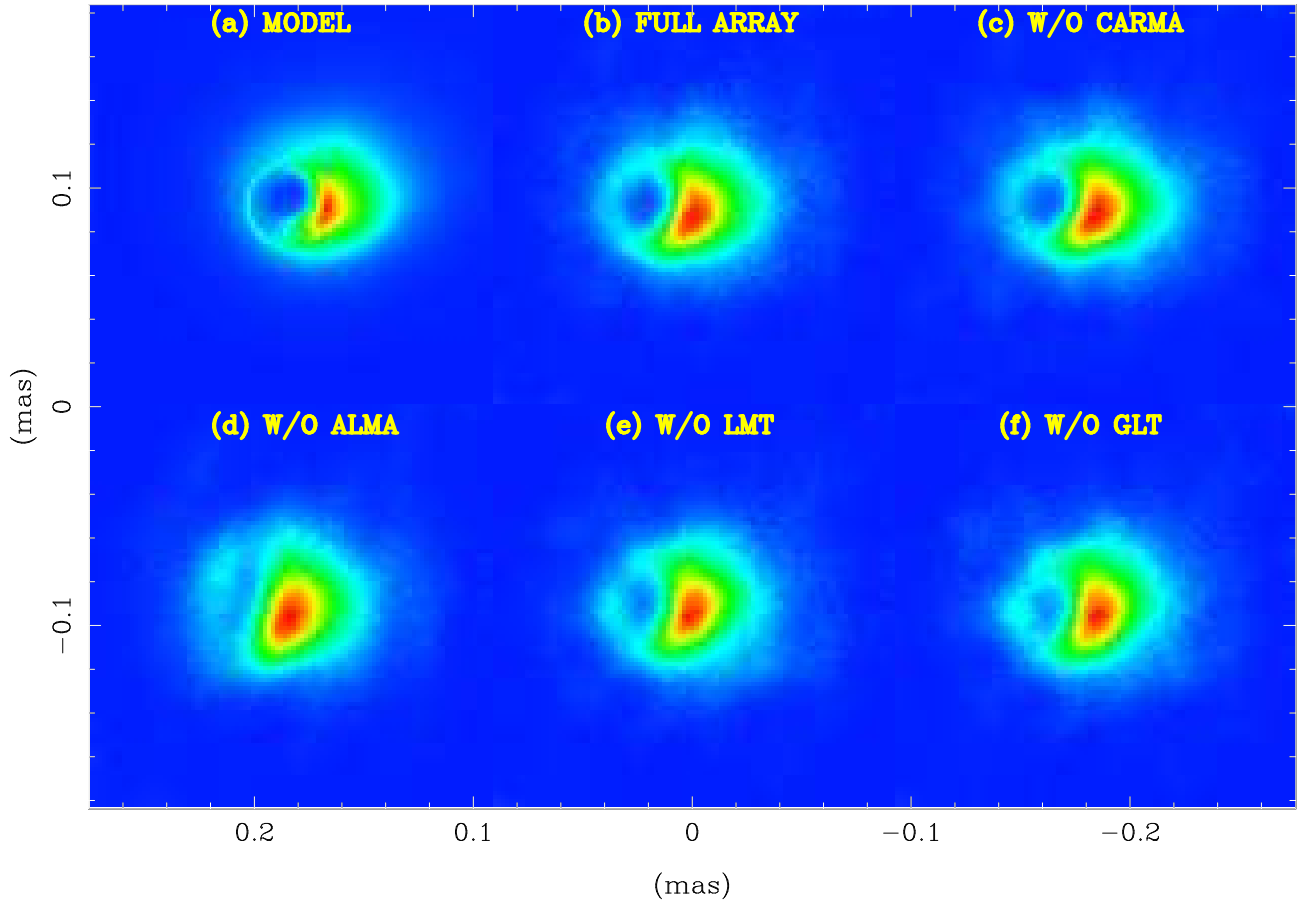}
 \caption{Degradation in image reconstruction at 230\,GHz. Panel (a) shows the input truth image, whereas the SQUEEZE reconstructions are shown in (b) with the full array, in (c) without CARMA, in (d) without ALMA, in (e) without LMT, and in (f) without GLT.}
 \label{Fig:degradation230}
 \end{center}
 \end{figure*}

Often a single site within the array is unavailable during a VLBI run due to, e.g., bad weather conditions or hardware failures. We consider here the image degradation compared to what is obtainable with the full array at 230\,GHz when a given site is unavailable, e.g., the phased ALMA, the phased CARMA, the LMT, and the GLT. We use the first model image in the sequence shown in Figure~\ref{Fig:230GHz_4G} as the input image.  Figure~\ref{Fig:degradation230} shows the truth image (panel a) and SQUEEZE reconstructed images with the full array (b), and the array without CARMA (c), the array without ALMA (d), the array without LMT (e), and the array without GLT (f).

A visual inspection indicates that the most severe degradation happens when the phased ALMA is missing (panel d). In this case, the shadow feature cannot be clearly imaged. This is because all the longest and most sensitive baselines are provided by ALMA  (Figure~\ref{Fig:visplt}). The degradation caused by dropping the CARMA, LMT and GLT leads to a slight decrease in MSE relative to that caused by dropping the phased ALMA (Table~\ref{Table:degradation230}). However, the DSSIM statistic does not confirm the same trend, as perceived by human observers.  In general, these pixel-based comparisons provide little understanding on how the morphology of black hole features differs from image to image. They might erroneously report pictures with similar pixel statistics to be similar, even though the black hole features look different to the human eye. Future development of feature-based metrics (i.e., metrics that characterize the morphological properties of black hole features) can potentially provide a more unbiased way for black hole image comparison. 

\begin{figure*}[ht!]
\begin{center}
\includegraphics[angle=-90,width=0.95\textwidth,clip]{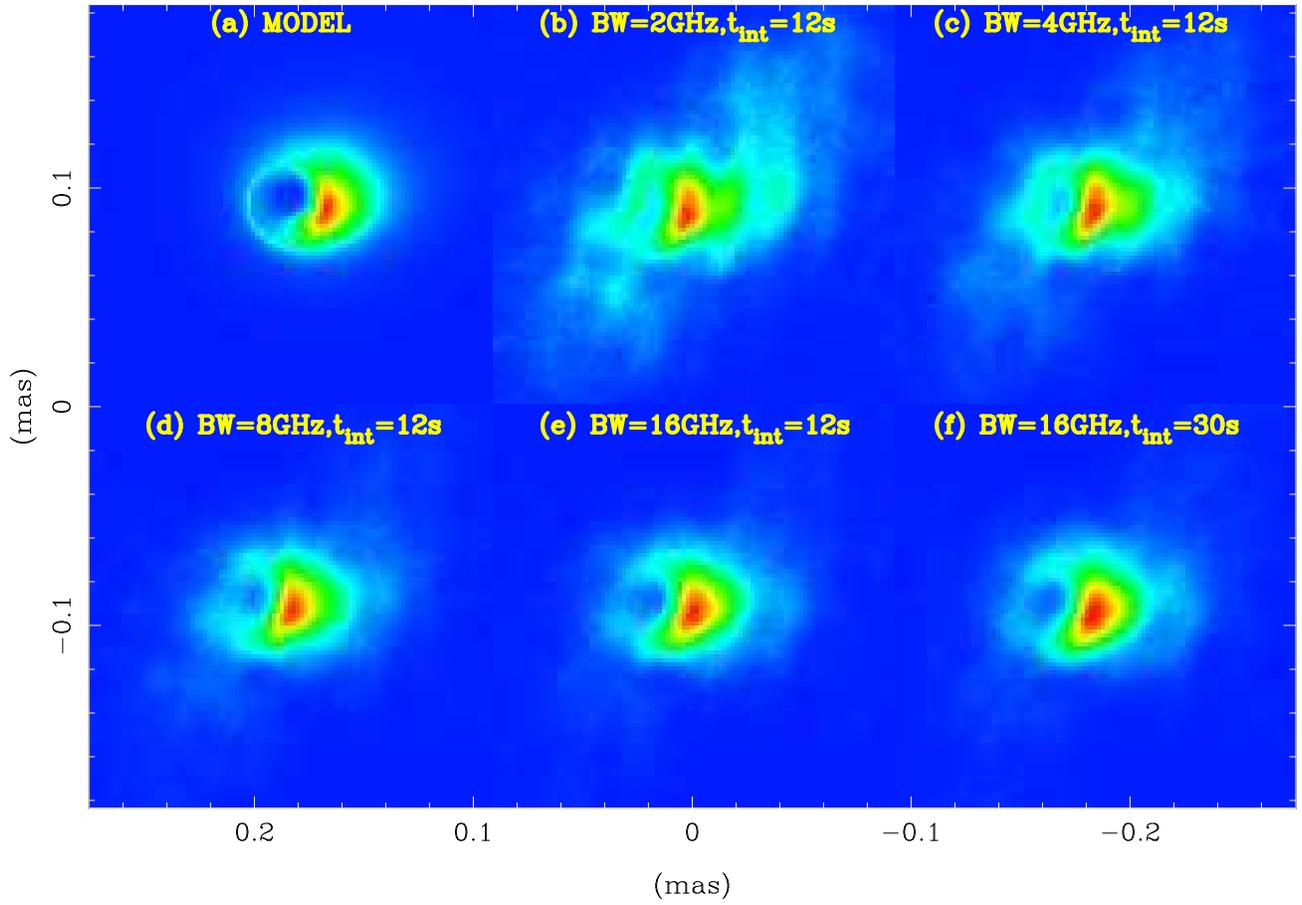}
\caption{A model image at 345\,GHz (a) and its SQUEEZE reconstructions assuming bandwidth from 2 to 16\,GHz by power of two from (b) to (e) and an integration time of 12\,s. (f) is same as (e), but with an integration time of 30\,s.}
\label{Fig:bw}
\end{center}
\end{figure*}

The array sensitivity may also differ from what we have assumed. This can result from weather conditions at individual sites, including varying atmospheric coherence times and phase stability among stations. Initial experiments at 345\,GHz may also be equipped with recording devices with data rates below the 64~Gbit~s$^{-1}$ assumed here, preventing observations that utilize the full 16 GHz bandwidth. We explore these effects by examining the simulated image reconstruction fidelity assuming different recording bandwidths from 2 to 16\,GHz by powers of two. Since it is likely that future radiometric phase compensation will be able to significantly increase the coherence time, we also consider longer integration times (30~s) at the highest data rate. Since the signal-to-noise ratio (S/N) for a coherently integrated signal improves as $\propto \sqrt{\Delta\nu t}$, here $\Delta\nu$ is bandwidth, bandwidth and coherence time equivalently improve the sensitivity. Figure~\ref{Fig:bw} shows the improvement in image quality as the array sensitivity increases, which is clearly presented by decreases in both the MSE and DSSIM metrics as shown in Table~\ref{Table:bw}. This indicates that we are limited mainly by the signal-to-noise ratio at 345\,GHz, rather than the limited sampling of the Fourier component for the considered array.

\begin{figure*}[ht!]
 \begin{center}
 \includegraphics[angle=-90,width=0.95\textwidth,clip]{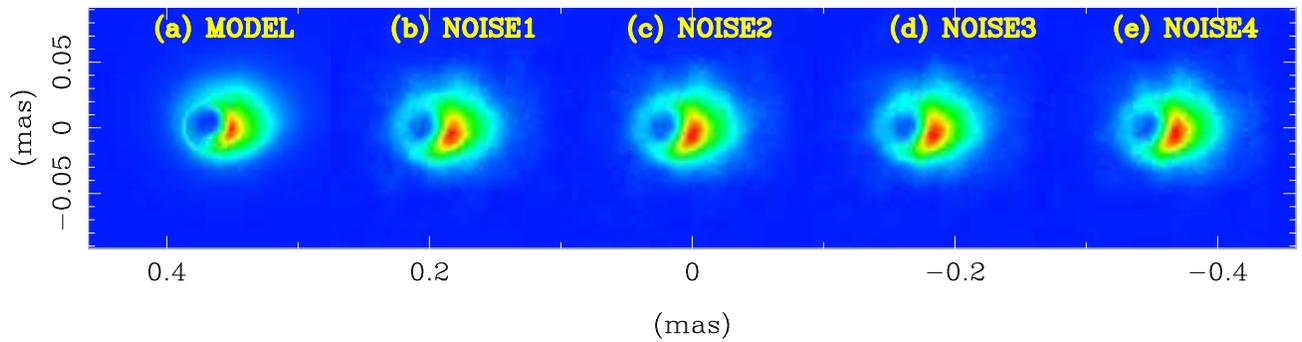}
 \caption{A model image at 230\,GHz (a) and its SQUEEZE reconstructions with four different realizations of the noise (from (b) to (e)).}
 \label{Fig:230GHz_noise}
 \end{center}
 \end{figure*}

Different noise realization in our simulation does not affect the quality of the reconstructed image and the shadow detectability. Figure~\ref{Fig:230GHz_noise} shows four reconstructions of a model image (first model image shown in Figure~\ref{Fig:230GHz_4G}) with the same assumed array characteristics but different noise realizations. We found a peak-to-peak change $\sim$ 7\,\% in MSE and 9\,\% in DSSIM, indicating none of the features are systematically affected.

 \begin{table}
\centering
\caption{Quality assessment for reconstructed images at 230\,GHz when a given site is unavailable~\label{Table:degradation230}.}
\begin{tabular}{cc|c}
\hline
Missing site&MSE&DSSIM\\
\hline
  -   &  0.019&  0.103\\
 CARMA&0.022&0.139\\
ALMA  & 0.037 & 0.132\\
LMT   & 0.032 & 0.227\\
GLT   & 0.030 & 0.124\\
\hline
\end{tabular}
\end{table}

 \begin{table}
\centering
\caption{Quality assessment for reconstructed images at 345\,GHz for a model image with various sensitivities~\label{Table:bw}.}
\begin{tabular}{ccc|c}
\hline
BW &t$_{int}$&MSE&DSSIM\\
(GHz)&(s)&- &-\\
\hline
2   &12 & 0.299 & 1.317\\
4   &12 & 0.155 & 0.618\\
8   &12 & 0.066 & 0.301\\
16  &12  & 0.038 & 0.180\\
16  &30  & 0.036 & 0.169\\
\hline
\end{tabular}
\end{table}

\section{SUMMARY}
\label{sect:summ}

M87 provides us with a unique opportunity for understanding and testing 
the black hole physics and astrophysical processes in relativistic jet formation, collimation, and propagation with the EHT.
With well-established algorithms tailored for the EHT, we have shown that the EHT is able to image the black hole shadow with limited data sampled in the Fourier domain. We have explored the dependence of jet structure upon the variation of the jet load radius and have shown that jet base activity in the vicinity of the black hole in M87 can be monitored with the EHT telescope. The expected structural variation time scale on these scales (a few $R_{sch}$) is well-matched with the duration of VHE flare activities. 

Our simulations indicate that minimum necessary requirements need to be met for clearly detecting the black hole shadow and studying event-horizon-scale jet launching in M87. The emission from the counter jet has to be sufficient for the black hole to cast a jet against, which in the context of the considered jet model corresponds to a load radius of $\lesssim$ 11 M. To obtain sufficient resolution and sensitivity, the phased ALMA has to be included in the array with bandwidth$\times$coherence time $\gtrsim$ 4\,GHz$\times$12 s at 230 GHz and more stringent sensitivity requirement at 345\,GHz. With the full array, the EHT in the next few years will be able to provide the best evidence for the presence of a black hole and to understand the jet-launching processes near the black hole in M87.

\acknowledgments
We thank the anonymous referee for suggestions that improved the quality of the paper. High frequency VLBI work at MIT Haystack Observatory is supported by grants from the National Science Foundation. This work is also supported through NSF 
grants AST-1310896, AST-1211539. We acknowledge support from the Gordon \& Betty Moore Foundation through award GMBF-3561. A.E.B.~receives financial support from Perimeter Institute for Theoretical Physics and the Natural Sciences and Engineering Research Council of Canada through a Discovery Grant. Research at Perimeter Institute is supported by the Government of Canada through Industry Canada and by the Province of Ontario through the Ministry of Research and Innovation. We thank Dr. Keiichi Asada for providing information for the Greenland telescope and Dr. Michael Johnson for useful comments.

 \end{document}